\documentclass[aps,prd,preprintnumbers,nofootinbib,twocolumn]{revtex4}
\usepackage{bm}
\usepackage{latexsym}
\usepackage{dcolumn}
\usepackage{amsmath,amsfonts,amssymb}
\usepackage{graphicx,epsfig}
\usepackage{color}
\usepackage{amsthm}

\def\be {\begin{equation}}
\def\ee {\end{equation}}
\def\bea {\begin{eqnarray}}
\def\eea {\end{eqnarray}}
\def\bc {\begin{center}}
\def\ec {\end{center}}
\def\bfg {\begin{figure}}
\def\efg {\end{figure}}
\def\bi {\begin{itemize}}
\def\ei {\end{itemize}}
\def\nn {\nonumber}

\def\le {\left}
\def\ri {\right}

\def\beq{\begin{equation}}
\def\eeq{\end{equation}}
\def\br{\begin{eqnarray}}
\def\er{\end{eqnarray}}
\newcommand{\eel}[1] {\label{#1}\end{equation}}

\newcommand{\bdm}{\begin{displaymath}}
\newcommand{\edm}{\end{displaymath}}

\begin{document}
\title{Nonsingular Rainbow Universes}

\author{Adel Awad $^{1,2}$} \email[]{aawad@zewailcity.edu.eg}
\author{Ahmed Farag Ali $^{1,3}$} \email[]{ahmed.ali@fsc.bu.edu.eg;afarag@zewailcity.edu.eg}
\author{Barun Majumder $^4$} \email[]{barunbasanta@iitgn.ac.in}

\affiliation{$^1$Center for Theoretical Physics, Zewail City of Science and Technology, Giza, 12588, Egypt.\\}
\affiliation{$^2$Dept. of Physics, Faculty of Science, Ain Shams University, Cairo, 11566, Egypt.\\}
\affiliation{$^3$Dept. of Physics, Faculty of Sciences, Benha University, Benha, 13518, Egypt.\\}
\affiliation{$^4$Indian Institute of Technology Gandhinagar, Ahmedabad, Gujarat 382424, India.\\}

\begin{abstract}
\par\noindent
In this work, we study FRW cosmologies in the context of gravity rainbow. We discuss the general conditions for having a nonsingular FRW cosmology in gravity rainbow. We propose that gravity rainbow functions can be fixed using two known modified dispersion relation (MDR), which have been proposed in literature. The first MDR was introduced by Amelino-Camelia, et el. in \cite{AmelinoCamelia:1997gz} and the second was introduced by Magueijo and Smolin in \cite{Magueijo:2001cr}. Studying these FRW-like cosmologies, after fixing the gravity rainbow functions, leads to nonsingular solutions which can be expressed in exact forms.
\end{abstract}

\maketitle


\section{Introduction}

Semi-classical approaches of quantum gravity (QG) are expected to play an important
role in revealing some essential features of the fundamental quantum theory of gravity.
One common feature among most of these semi-classical approaches
\cite{guppapers,BHGUP,Scardigli} is the existence of a minimal observable
length $l_p$, i.e. Planck length. This minimal length works as a natural
cutoff, which is expected to resolve the known curvature singularities in
general relativity. Another feature, in some of these approaches \cite{kmm,kempf,brau},
is the departure from the relativistic dispersion relation by redefining the
physical momentum at the Planck scale, or Lorentz invariance
violation. This departure could also be a result of spacetime \cite{'tHooft:1996uc}
discreteness, spontaneous symmetry breaking of Lorentz invariance in string field
theory\cite{LIstring}, spacetime foam models \cite{AmelinoCamelia:1997gz} or
spin-network in Loop quantum gravity (LQG) \cite{Gambini:1998it}. Besides,
there are other approaches such as non-commutative geometry \cite{Carroll:2001ws}
which predicts a Lorentz invariance violation. These studies indicate that
Lorentz violation could be a theoretical possibility in several approaches
investigating QG. These approaches predict a departure from Lorentz invariance
in the form of modified dispersion relations (MDR). It has been
argued that MDR could explain the threshold anomalies occurring in ultra
high energy cosmic rays and TeV photons \cite{AmelinoCamelia:1997gz,AmelinoCamelia:1997jx,AmelinoCamelia:1999wk,s8}. For a recent detailed review along the mentioned lines can be found in \cite{amerev}.\par
One of the most interesting forms of MDR, has been suggested by Amelino-Camelia, et el. in
\cite{AmelinoCamelia:1997gz,AmelinoCamelia:1997jx,AmelinoCamelia:1999wk} which has the following form;

\be
p^2= E^2 \le(\frac{e^{E/E_{Pl}}-1}{E/E_{Pl}}\ri)^2 \label{MDR}
\ee

where $E_{Pl}$ describes the energy scale at which the dispersion relation is modified and it
is taken to be the Planck energy. This modified dispersion relation has been introduced by Amelino-Camelia, et al. to explain the astrophysical observations of the hard spectra coming from gamma-ray bursters \cite{AmelinoCamelia:1997gz,AmelinoCamelia:1997jx,AmelinoCamelia:1999wk} at cosmological distances.\par
A theoretical frame work that naturally produce MDR is double special relativity (DSR)\cite{AmelinoCamelia:2000mn}. DSR is an extension of special relativity which preserves the relativity principle and extends the invariant quantities to be the Planck energy scale beside the
speed of light. The simplest realizations of the idea of DSR \footnote{It is worth mentioning that DSR is one possibility to address the issue of
 constructing a non-linear Lorentz transformation in momentum space but there are other alternatives also. Some recent developments include \cite{refadd1}.} are based on a non-linear Lorentz transformation in momentum space, which imply a deformed Lorentz symmetry such that
the usual dispersion relations in special relativity may be modified by Planck scale corrections.
It should be mentioned that Lorentz invariance violation and Lorentz invariance
deformation are in general conceptually different scenarios. Here we are going to adopt Lorentz invariance
deformation scenarios by considering DSR and its extension in models of rainbow gravity.

In the framework of DSR the definition of
the dual position space is not trivial due to the nonlinearity
of the Lorentz transformation. To overcome this issue,
Magueijo and Smolin \cite{Magueijo:2002xx} proposed a doubly general relativity which
assumes that the spacetime background felt by a test particle would depend on its energy. Therefore, we will not have a single metric describing spacetime, but a one parameter family of metrics which depends on the energy (momentum) of these test particles, forming
a {\it rainbow} geometry. This approach is known as \emph{Gravity Rainbow} and can be understood as follows; the non-linear of Lorentz transformation leads to
the following modified dispersion relation
\be
E^2 f(E/E_{Pl})^2- p^2 g(E/E_{Pl})^2= m^2 \label{RainbowDisper}
\ee

where $E_{Pl}$ is the Planck energy scale, m is the mass of the test particle,
 $f(E/E_{Pl})$ and $g(E/E_{Pl})$ are commonly known as Rainbow functions and $\lim_{E\rightarrow 0} f(E/E_P) =1$
 and $\lim_{E\rightarrow 0} g(E/E_P) =1$.\par
A modified equivalence principle was proposed in \cite{Magueijo:2002xx} which requires that one parameter family of energy dependent orthonormal frame fields describe a one parameter family of energy dependent metrics given by
\begin{equation}
h(E/E_P) = \eta^{ab} e_a(E/E_{Pl}) \otimes e_b(E/E_{Pl})
\end{equation}
where $e_0(E/E_{Pl}) =(1/f(E/E_{Pl})) \tilde{e}_0$ and $e_i(E/E_{Pl})= (1/f(E/E_{Pl})) \tilde{e}_i$. But in the limit $(E/E_{Pl})\rightarrow 0$ general relativity must be recovered. With the definition of one parameter
family of energy momentum tensors Einstein's equations are also modified as
\begin{equation}
G_{\mu \nu}(E/E_{Pl}) = 8\pi G T_{\mu \nu}(E/E_{Pl}) \label{RFE}
\end{equation}
Potential investigations on the gravity rainbow can be found in \cite{Galan:2004st}.\par
The choice of the Rainbow functions $f(E/E_{Pl})$ and $g(E/E_{Pl})$ is very important for making predictions. Among different arbitrary choices in \cite{Galan:2004st,FRWRainbow}, many aspects of the theory have been studied with Schwarzschild metric, FRW universe and black hole thermodynamics. In this letter we employ the modified dispersion relation of Eq.(\ref{MDR}) \cite{AmelinoCamelia:1997gz}, which fix the rainbow functions
$f(E/E_{Pl})$ and $g(E/E_{Pl})$. We also employ another choice of rainbow functions $f(E)=g(E)=\frac{1}{1-E/E_{Pl}}$ which was considered in \cite{Magueijo:2002xx,Magueijo:2001cr}. This particular choice is capable of giving a theory with constant velocity of light and also solves the horizon problem. Using these rainbow functions we study the effect of gravity rainbow,
on the FRW universe and investigate the new properties of FRW universe
in the existence of the MDR.\par

\section{FRW Rainbow Cosmology}

Here we review FRW universe in rainbow gravity and study its effect as a semi-classical approach of QG in the early universe\cite{FRWRainbow}. The most general FRW universe in rainbow gravity has been found in \cite{FRWRainbow}, it has the following metric
\be
ds^2=-\frac{1}{f(E)^2}dt^2+\frac{a^2}{g(E)^2}dx^2 ~~,
\ee

where we have considered a spatially flat universe (i.e., $k=0$). Using the above metric the authors in \cite{FRWRainbow} found the following modified Friedmann equations:

\bea
\le(H-\frac{\dot{g}(E)}{g(E)}\ri)^2&=&\frac{8 \pi G}{3 f(E)^2} \rho  \label{FR1}\\
\dot{H}+\frac{\dot{g}(E)^2}{g(E)^2}-\frac{\ddot{g}(E)}{g(E)}&=&-\frac{4 \pi G (\rho+P)}{f(E)^2}\nn\\&&-\le(H-\frac{\dot{g}(E)}{g(E)}\ri)\frac{\dot{f}(E)}{f(E)}\label{FR2}
\eea
where $H= \frac{\dot{a}}{a}$. In addition, the conservation equation is modified to

\be
\dot{\rho}+3(H-\frac{\dot{g}}{g})(\rho+P)=0 \label{Conserv}
\ee
Here we adopt the point of view of Ref.\cite{FRWRainbow} to study the effect of rainbow gravity as a semi-classical approach of QG in the early universe. As we have mentioned earlier the geometry probed by a test particle depends on its energy. This raises the question; which metric one might use to describe the evolution of the spacetime. In this framework we consider a large ensemble of ultra relativistic particles (dominant in the early universe) which are in thermal equilibrium and has a typical or an average energy $\epsilon\sim T$. As in standard cosmology the continuity equation leads to the first law of thermodynamics \be d(\rho\, V)=-P\, dV,\ee where $V=(a/g)^3$. The above equation and integrability condition $ {\partial^2 S \over \partial V \partial P} ={\partial^2 S \over \partial P \partial V} $  \cite{Kolb:1990vq} leads to a constant entropy
\be S= {V(\rho+P) \over T}=const.\ee In this section and the coming sections we are going to take the pressure to be $P=(\gamma-1) \,\rho$, as in standard cosmology which leave the FRW spacetime singular at $t=0$. For such a pressure, the average energy $\epsilon$ can be expressed as \be \epsilon \sim T = c'\, \gamma\, V \rho,\ee where $c'$ is some constant.
Using the above equation of state (EoS) in the conservation equation Eq. (\ref{Conserv}) we get the following equation
\be {d\rho \over d\ln(a/g)}=-3\, \gamma\, \rho\ee
which can be solved to give a density  $\rho \propto (a/g)^{-3\gamma}$. This leads to an average energy
 \be \epsilon = c\, \gamma\, \rho^{{\gamma-1} \over \gamma}.\label{aenergy}\ee
Notice that the above relation between the average energy $\epsilon$ and the density $\rho$ depends only on the EoS parameter $\gamma$. Clearly, the above relation does not depend on the form of MDR chosen for a particular model.

\section{ When a Nonsingular Rainbow Universe Possible}

Before showing how the MDR mentioned above leads to a nonsingular FRW-like metric, it is constructive to discuss the general conditions on the rainbow functions that leads to a nonsingular cosmology. Substituting the modified Friedmann equation Eq.(\ref{FR1}) in the continuity equation Eq.(\ref{Conserv}) one gets

\bea
\dot{\rho}=-\sqrt{24 \pi G} \,\gamma \,{\rho^{3/2} \over f(\rho)}=K(\rho) , \label{Conserv11}
\eea
where we wrote $f$ as a function of $\rho$ instead of $\epsilon$ according to Eq.(\ref{aenergy}) and define some function $K(\rho)$. This first-order system is well studied in dynamical system (see  e.g., \cite{Awad:2013tha}, or see \cite{strogatz} for more general applications) in cosmological contexts. Knowing the fixed points of the function $K(\rho)$, (i.e., its zeros, let us call them $\rho_i$) and its asymptotic behavior enables one to qualitatively describe the behavior of the general solution without actually solving the system. Fixed points are classified according to their stability to stable, unstable, or half-stable. In \cite{Awad:2013tha} a very similar system has been studied which was expressed in terms of the Hubble rate. It is straight forward to use the same analysis to study the density $\rho$ instead of the Hubble rate $H$.

Our basic idea for resolving finite-time singularities is to show the existence of an upper bound for the density $\rho$ (through having a fixed point $\rho_1$) which is reached at an infinite time, or to show the existence of a point at which the density is unbounded (a potential singularity) but reached in an infinite time, i.e., not a physical singularity. Therefore, following the discussion in \cite{Awad:2013tha}, one can show that finite-time singularities (including big bang singularities) are absent if one of the following is true; i) If $f$ grows asymptotically as $\sqrt{\rho}$, or faster. For example, if $f\sim \rho^{s}$, where $s\geq 1/2$. In this case, one can calculate the time to reach a potential singularity by integrating Eq. (\ref{Conserv11}) starting from some initial finite density $\rho^*$ to an infinite one. This integral leads to \be t= c'' \int_{\rho^*}^{\infty}\rho^{(s-3/2)}\, d\rho=\infty\hspace{.23in}, s\geq 1/2 \ee This means that the time to reach this potential singularity is infinite, therefore, it is not a finite-time singularity, i.e., not physical. ii) If $f^{-1}$ is differentiable and has a zero at $\rho=\rho_1$ (notice that in this case the function $K(\rho)$ will have two fixed points, namely, $0$ and $\rho_1$), then according to the analysis in \cite{Awad:2013tha} the cosmological solution is nonsingular and interpolates monotonically between $\rho_1$ and $0$.

With a particular choice of $f(E)$ and $g(E)$ a possible resolution of the big bang singularity was proposed in \cite{bar} in the context of quantum cosmology with a perfect fluid. The rainbow function $f(E)$ plays an important role in possible resolution of the big bang singularity, but it has to satisfy one of the above conditions to do that.

We conclude that FRW-like rainbow cosmologies are nonsingular (expressing $f$ as a function of the energy $E$ after using Eq.\ref{aenergy}) if at least one of the following is true;\\
i) $f$ grows asymptotically as $E^{\gamma \over 2 (\gamma-1)}$, or faster,\\
ii) $f^{-1}$ is differentiable and has a zero at $\rho_1\neq 0$.\\

Although the following two sections we are not going to discuss MDR's with power law behavior and integer exponents, it is worth mentioning that these MDR's are widely studied for phenomenological purposes, see for example \cite{AmelinoCamelia:1997gz,AmelinoCamelia:2000mn}. For this reason, we would like to apply the above criteria on these cases and comment on that. Assuming that the asymptotic behavior of the rainbow function takes the form $f \sim E^n$, where $n$ is an integer. According to the previous paragraph, the cosmological solution is nonsingular when $f$ grows as $E^{\gamma \over 2 (\gamma-1)}$ or faster. This leads to an inequality between the parameters "$n$" and "$\gamma$" which reads $n \geq {\gamma \over 2 (\gamma-1)}$. For example, taking $\gamma=4/3$, leads to a nonsingular solution if $n \geq 2$, i.e., if $f$ grows as $f \sim E^2$ or faster. Notice that, the cases which doe not satisfy the previous criteria can still have nonsingular solutions if they satisfy criteria $ii)$. Therefore, one is lead to the following comment; among the cases where $f \sim E^n$, the most interesting ones in cosmological contexts are those with $n\geq 2$, since they are free from finite-time singularities.

In this work we present two interesting MDR's which lead to nonsingular FRW-like cosmologies. The first belongs to case $i)$, and the second belongs to case $ii)$ as we will see in the coming sections.\par

\section{Nonsingular Rainbow Universes}
Now we employe the modified dispersion relation which is proposed by the Amelino-Camelia, et al. in Eq. (\ref{MDR}),
and compare it with Eq. (\ref{RainbowDisper}). The functions $f(E)$
and $g(E)$ can be fixed as follows:

\bea
f(E)=\frac{e^{E/E_{Pl}}-1}{E/E_{Pl}},~~~~~~~~g(E)=1. \label{RainFunc}
\eea

Using this identification, and Eq. (\ref{aenergy}), the function $f(\epsilon)$ will be:

\bea
f(\epsilon)=\frac{\exp{(\gamma \, {\varrho}^{\frac{\gamma-1}{\gamma}})}-1}{ \gamma\, {\varrho}^{\frac{\gamma-1}{\gamma}}}
\eea
where $\varrho =\rho / \rho_p$ and wrote the Planck energy in terms of some density $\rho_p$ as $E_{Pl}=c\, {\rho_p}^{{\gamma-1 \over \gamma}}$. Here we are
interested in an EoS parameter $ \gamma > 1$, particularly the case where $\gamma = 4/3$ (i.e., radiation), but we are going to leave the expressions as general as possible.

\begin{figure}
\includegraphics[scale=0.35,angle=270]{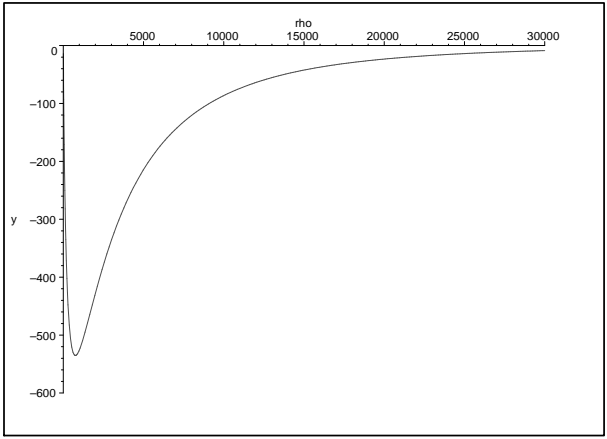}
\caption{$\dot{\rho}$ versus $\rho$}.
\label{solution}
\end{figure}

The modified Freidmann equation due to MDR will be given as follows:

\bea
H^2&=&\frac{8 \pi \,{\rho_p}^{\frac{2-\gamma}{\gamma}}\,{\varrho}}{3} \le(~~\frac{\gamma\,{\varrho}^{\frac{\gamma-1}{\gamma}}}{{\exp{({\gamma\, {\varrho}^{{\frac
{\gamma-1}{\gamma}}}}})}-1}~~\ri)^2.  \label{FR3}
\eea
Here, we investigate a possible resolution of the big bang singularity using the discussion in section $III$ (see \cite{Awad:2013tha} for more detailed analysis). Using the modified Friedmann equation of Eq. (\ref{FR3}) in Eq. (\ref{Conserv}), we get the following equation

\be
\dot{\varrho}=-\sqrt{24 \pi}\,\gamma^2\,{\rho_p}^{\frac{2-\gamma}{2\gamma}}\, \le(~~\frac{{\varrho}^{\frac{5\gamma-2}{2\gamma}}}{{\exp{({\gamma\, {\varrho}^{{\frac
{\gamma-1}{\gamma}}}}})}-1}~~\ri) . \label{Conserv1}
\ee
This put the continuity equation in the form $\dot{\rho}= K(\rho)$, which is useful in describing the behavior of the general solution without having the form of the exact solution. One can observe that the above MDR might be able to resolve the big bang singularity since $f$ grows asymptotically faster than $\sqrt{\rho}$.  To see that let us first plot $\dot{\rho}$ versus $\rho$ in Fig. \ref{solution}, where we consider the relativistic case $\gamma=4/3$ and $\rho_p=1$. From the plot or simple analysis one can observe that the density is not bound which might be a sign of singularity.

Now let us show that the time taken by the solution to evolve from some finite density $\varrho^*$ to an infinite one (a potential singularity) is infinite. This can be done by integrating Eq. (\ref{Conserv1})
\be
t= -\int_{\infty}^{\varrho^{\star}}\, \frac{9}{32\sqrt{6\pi}}\,\rho_p^{-\frac{1}{4}}\,{\varrho^{-\frac{7}{4}}}{[\,\exp{({{3/ 4}}\,\varrho^{\frac{1}{4}})}-1\,]}d\varrho=\infty
\ee
Since the time to reach this infinite density is infinite, there is no finite-time singularities. Similarly, the solution takes an infinite time to reach the fixed point $\rho=0$ starting from a finite value $\varrho^{\star}$, this can be calculated as
\be
t= -\int_{\varrho^{\star}}^{0}\, \frac{9}{32\sqrt{6\pi}}\,\rho_p^{-\frac{1}{4}}\,{\varrho^{-\frac{7}{4}}}{[\,\exp{(\frac{3}{4}\,\varrho^{\frac{1}{4}})}-1\,]}d\varrho=\infty
\ee

Considering the above time calculations, one can see that the solution does not suffer from any finite-time singularity. Notice that this nonsingular behavior of the modified continuity equation is
valid for all $\gamma > 1 $.

\begin{figure}
\includegraphics[scale=0.35,angle=270]{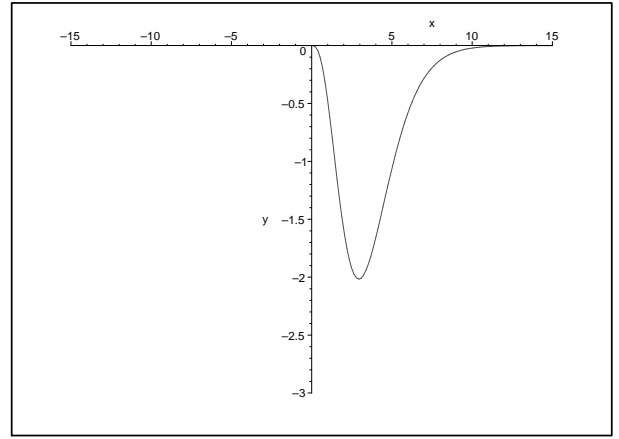}
\caption{$\dot{x}$ versus $x$}.
\label{solution2}
\end{figure}

Let us try to find an exact solution for Eq. \ref{Conserv1} in terms a new variable $x$, where $x=\frac{4}{3}\,\varrho^{1/4}$ and a rescaled time $\tau=\sqrt{\frac{27}{2}\pi}\, {\rho_p}^{1/4}$. Clearly $x$ is inversely proportional to the scale factor, or $x\sim a^{-1}$. Now Eq. \ref{Conserv1} becomes
\be \frac{dx}{d\tau}=-\frac{1}{4}\,x^4\,e^{-x}\ee
which can give an exact expression relating the variable $x(t)$ and the time $\tau$
\bea
\tau=&& \frac{2}{3}\,{ e}^{x} \le(\frac{2}{x^3}+\frac{1}{x^2}+\frac{1}{x}\ri)+\frac{2}{3}\,Ei(1,-x)+C,
\eea
where $Ei$ is the exponential integral. Another way of showing the behavior $x$ as function of time is to plot $\dot{x}$ versus $x$ which produce the graph in Fig. \ref{solution2}.

Finally, it is important to check the  behavior  the MDR of Eq.(\ref{MDR}) at the Planck scale to investigate if density of states
would diverge or not \cite{FRWRainbow}. We find that the density of states are defined as follows

\bea
\mathcal{G}(E) dE &\simeq& 4 \pi p^2 dp\nn\\
&\simeq& f(E)^3 \le(1+ E \frac{f(E)^{\prime}}{f(E)}\ri) E^2 dE
\eea

By substituting with the form of $f(E)$ from Eq.(\ref{MDR}), we find that the density of states
has a finite value $e (e-1)^2$ and does not have any divergence behavior.

\section{Magueijo-Smolin dispersion relation}

In this section, we use the dispersion relation proposed by Magueijo-Smolin in \cite{Magueijo:2001cr} to fix the rainbow functions and investigate
its impact on FRW universe through the gravity rainbow approach \cite{Magueijo:2002xx}. According to the Magueijo-Smolin dispersion relation \cite{Magueijo:2001cr}, the rainbow functions can be fixed as follows:

\bea
f(E)=g(E)=\frac{1}{1-E/E_{Pl}}
\eea

Using the same steps in the previous section one can express $f$ and $g$ in terms of $\rho$ as follows:
\begin{equation}
\label{fgrho}
f(\rho) = g(\rho) = \le({1-\frac{\gamma\, \rho^{\frac{\gamma-1}{\gamma}}}{E_{Pl}}}\ri)^{-1}
\end{equation}

\begin{figure}
\includegraphics[scale=0.35,angle=270]{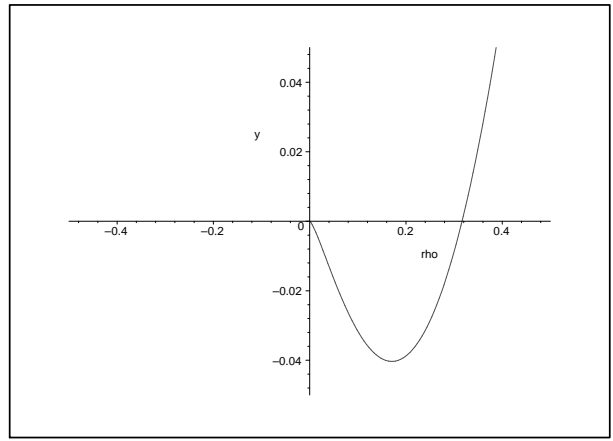}
\caption{$\dot{\rho}$ versus $\rho$}.
\label{solution3}
\end{figure}

By expressing the modified continuity equation \ref{Conserv11} in terms of $\varrho$ one gets the following expression

\begin{equation}
\label{dotrho}
\dot{\varrho} = -~\sqrt{{24\pi}} ~\gamma ~{\rho_p}^{\frac{2-\gamma}{2\gamma}}\varrho^{3/2} \left(1 - \gamma\, \varrho^{\frac{\gamma-1}{\gamma}}\right)
\end{equation}
Now following the discussion in section III, one can observe that the above system has two fixed points, $\varrho=0$ and $\varrho=\gamma^{\gamma \over 1-\gamma}$, which is showing that the solution is nonsingular and interpolate between $\rho=0$ and $\rho\sim \rho_p$. Let us consider the relativistic case $\gamma=4/3$, where the relation between $\dot{\rho}$ and $\rho$ is depicted in Fig. (\ref{solution3}). One can show the absence of finite-time singularities by calculating the time necessary to reach any of the two fixed point $\rho_f=0$ or $\rho_f= ({3 \over 4})^4 \rho_p$ (starting from a finite density $\rho^{\star}$)
\be
t= -\int_{\varrho^{\star}}^{\rho_f}\, \frac{3}{8\sqrt{6\pi}}\,\rho_p^{-\frac{1}{4}}\,\varrho^{-\frac{3}{2}}\,[1- 4/3 \varrho^{1/4}]d\varrho=\infty,
\ee
which means that the time necessary to reach a fixed point is infinite. This introduce
a possible resolution for the big bang singularity.
This gives a solution which is non-singular and has two fixed points.
To calculate the time which is necessary to reach these fixed points, we should calculate
$\rho$ as a function of time.

We find that Eq. (\ref{dotrho}) has an exact solution for the relativistic case $\gamma=4/3$. To show this let us write the above Eq. in terms of a new variable $y$, where $y=\frac{3}{4}\,\varrho^{-1/4}$ and a rescaled time $\tau'=\sqrt{\frac{27}{32}\pi}\, {\rho_p}^{1/4}$. Notice that, $y$ is proportional to the scale factor, or $y\sim a$. Now Eq. \ref{Conserv1} becomes

\be \frac{dy}{d\tau'}=\,y^{-2}\,(y-1)\ee

\be
\tau'=\frac{1}{2}\,y^2+y+\ln{(y-1)}+C.
\ee

\section{Conclusions}

In this work we studied the effect of rainbow gravity\cite{Magueijo:2002xx,FRWRainbow} as a semi-classical approach of quantum gravity in the early universe. We discussed the general conditions for having a nonsingular FRW cosmology in gravity rainbow. Fixing the rainbow functions using Amelino-Camelia, et al.\cite{AmelinoCamelia:1997gz} and Magueijo-Smolin in \cite{Magueijo:2001cr} enabled us to investigate its impact on FRW-like cosmology through the gravity rainbow approach. Studying the general cosmological solutions in both cases reveals the absences of big bang singularities and both solutions are nonsingular. Furthermore, these nonsingular solutions can be expressed in exact forms. The Friedmann equations are modified in the rainbow gravity formalism by the so called rainbow functions. In the first case, we have identified the rainbow functions with the modified dispersion relations as introduced by Amelino-Camelia, et al. and Magueijo-Smolin and studied the rainbow modified Friedmann equations with a perfect fluid. The conservation equation allowed us to evaluate a relation between the test particle energy and the energy density which is used to evaluate the modified continuity equation whose solutions played a major role in studying the singularity. For the modified dispersion relation as proposed by Amelino-Camelia, et al. we found non-singular solutions for a wide range of values for the equation of state parameter $\gamma > 1$. Using the analysis in \cite{Awad:2013tha} we notice that the universe takes infinite amount of time to reach $\rho \rightarrow \infty$ and $\rho=0$ from a finite value of $\rho$. We have also found that the density of states do not diverge at the Planck scale. In the second case, we used the MDR of Magueijo and Smolin and did the same analysis in \cite{Awad:2013tha} we find the system exhibits two fixed points, one of them is around the Planck scale. Also the system takes infinite time to reach the fixed points which represents a non-singular solution. So in both the cases we find a possible resolution of the big bang singularity.

\subsection*{Acknowledgments}
The research of AFA is supported by Benha University and CTP in Zewail City. BM was partly supported by the Excellence-in-Research Fellowship of IIT Gandhinagar. The authors would like to thank an anonymous referee for enlightening comments and helpful suggestions.



\end{document}